\documentstyle[preprint,aps]{revtex}
\begin{document}
\draft
\title{Ising spin glass with arbitrary spin beyond the mean field theory}
\author{\rm K. Walasek, K. Lukierska- Walasek, and M. Wodawski}
\address{ Institute of Physics, 
Pedagogical University, Plac
S{\l}owia\'nski 6,\\ 65- 069 Zielona G\'ora, Poland.} 
\maketitle
\begin{abstract}
We consider the Ising spin glass for the arbitrary spin $S$ with the short- 
ranged interaction using the Bethe- Peierls approximation previously
formulated by Serva and Paladin for the same system but limited to
$S=1/2$. Results obtained by us for arbitrary $S$ are not a simple
generalization of those for $S=1/2$. In this paper we mainly 
concentrate our studies
on the calcutation of the citical temperature and the linear
susceptibility in the paramagnetic phase as functions of the
dimension of the system and spin number $S$. These dependences are
illustrated by corresponding plots.   
\end{abstract}
\pacs{75.10.Nr, 75.50.Lk}
\section{Introduction}\label{sec:intro}
The study of glasses is today one of the most relevant and actual
problem in condensed matter physics. In particular research around
spin glasses in finite dimensions is very active since it is still
unclear if they share some the qualitative features of the mean-
field theory of the model Sherrington- Kirkpatrick (SK)\cite{1,2}
However, there are recent studies
\cite{3a,3,4} which indicate difficulties to extend the 
molecular field approximation (MFA) scenario to realistic spin glasses with 
short- range
interaction and decide "a priori" which properties survive and which
must be appropriately modified.  
 Recently, in an interesting paper \cite{5}, an approach
beyond the MFA has been achieved for an $d$- dimensional Ising SG
model ($S= 1/2$) with short- range interactions on a real lattice using an
extension of the Bethe- Peierls approximation (BPA) \cite{5} to the
spin glass problem via the replica trick. This approach seems to be
very promising to estabilish a direct contact with the results
obtained by different authors for the infinite- ranged version and to
controll possible deviations for short- ranged glasses from the well
acquired MFA scenario. Quite recently \cite{7} the Parisi \cite{2} ansatz has
been investigated for the Ising spin glass with $S= 1/2$ using the
generalized of the Bethe- Peierls metod named by the
authors "a variational approach" where finite clusters of spins
interacts and the sample averaging is properly taken into account.
The result is qualitatively similar to that obtained in the frame of
the MFA with some quantitatively
modifications due to short- range order interactions.

All studies mentioned above has been performed for the standard Ising
model with $S=1/2$. Therefore it seems to be quite interesting to extend the 
methods applied for the Ising spin glass, where the number of spin is
arbitrary.  In the spite of numerous works on the Ising model $S=1/2$ a 
little attention has been devoted to the same system with $S$ 
arbitrary. In the spin glasses solutions for the
arbitrary $S$ is 
not a simple generalization of that for the Ising model with $S=1/2$.
The  reason is that for higher spins
$\left(S_i\right)^n_i\neq S_i$ or $const$ which leads
to parameters diagonal in replica indices. This considerable affects
the results for the $S=1/2$ Ising SG.  

We will
use the Bethe- Peierls approximation (BPA) \cite{5} with some necessary 
modifications. As we will
see the critical temperature depends as well as of the dimension of the
system and of the spin number. In addition, linear
susceptibility even in the paramagnetic phase are nontrivial 
functions of the temperature.

The paper is planned as follows: In Section \ref{sec:BPA}  we shortly present the 
idea of the Bethe- Peierls approach for the the Ising spin glass \cite{5}
and  give equations for the critical temperature  and the linear 
susceptibility in the paramagnetic phase when the spin number $S$ is arbitrary.
 The Section
\ref{sec:res} contains 
plots as the critical temperature depends of the dimension and the
value of the spin number. Finally in Section  \ref {sec:final} some conclusion are 
drawn.

\section{Bethe- Peierls method for spin glasses}\label{sec:BPA}
The our starting point is the Hamiltonian of the Ising model with the
arbitrary spin:
\begin{equation} \label{1} H=-\frac{1}{2}\Sigma_{i,j} J_{i,j}S_iS_j\;\;,
\end{equation} where $S_i=S_i^z$ is the $z$-component of the spin.
As usual eigenvalues of $S_i$ runs from $-S$ to $S$, where $S$ is
arbitrary. In (\ref{1}) the summation over $i,j$ comprises only nearest
neighbours. Interaction parametrs $J_{i,j}$ are variables
obeying random distribution that is , dichotomic, gausssian, etc. For the
simplicity, in order do not complicated the main idea, we will assume
that $J_{i,j}$ is a random variable with $J_{i,j}=\pm J$ with the
equal probability for $+$ and $-$ sign.

Using the replica trick the free energy can be written as follows:
\begin{equation} \label{2} -\beta F=\rm{lim}_{n
\rightarrow}\frac{1}{n}\rm{Tr}\exp\left[-\beta\sum_{\alpha}
H_{\alpha}\right]_{\rm{av}}\;\;, \end{equation} where $\left[\cdots
\right]_{av}$ denotes a sample averaging and $H_\alpha$ is the
replicated Hamiltonian (\ref{1}).

Working directly on the real lattice, the
basic idea of the BPA for spin glasses \cite{5} is to take into
account the correct interactions inside replicated clusters (cl),
constitued by a central spin $ S_0$ and its $2d$ nearest-
neighbours $ S_i;\; \{i=1,\cdots ,2d\}$, and to describe the
interactions of the cluster borders with the remnant of the
system by means of effective couplings among replicas to be
determined self - consistently. Therefore the free energy in this
approximation can be written as follows:    
\begin{equation}\label{3} -\beta F= \text{lim}_{n 
\rightarrow 0}\frac{1}{n}\ln Z_n \;\;,
\end{equation} where \begin{eqnarray}\label{4} Z_n &=&K\left(\beta\right) 
\text{Tr}\Biggl\{\left[\exp\left(\beta\sum_{i=1,\alpha}^{2d}J_{0,i}S_{0,\alpha}
S_{i,\alpha}
\right)\right]_{\text{av}}\exp\left(
\frac{\beta^2J^2}{2}\mu\sum_{i=1}^{2d}\sum_\alpha S_{i,\alpha}^2\right)\\ \nonumber &\times
&\exp\left(\frac{\beta^2J^2}{2}\sum_{i=1}^{2d}\sum_{\alpha\neq\alpha^{\prime}}
\mu_{\alpha,\alpha^{\prime}}
S_{i,\alpha}S_{i,\alpha^{\prime}}\right)\Biggl\}\;\;.
\end{eqnarray}
In Eq. (\ref{4}) $\alpha=1\cdots n$ denotes the replica indices, $K\left(\beta
\right)$ is a multiplicative constant
which depends on the temperature but not on the lateral spins of the
cluster, $\mu_{\alpha,\alpha^{\prime}}$ and
$\mu\equiv\mu_{\alpha,\alpha}$, according to Bethe- Peierls ansatz,
describe the interaction betwen the "external world" and the lateral
spins of the replicated cluster. The difference between Eq. (\ref{4})
and the corresponding formula of Ref. \onlinecite{5} is that in
(\ref{4}) we have the additional parameter $\mu$. This is a consequence of the
fact that for an arbitrary spin $S_{k,\alpha}^2\neq 1/4$.

Effective couplings $\mu_{\alpha,\alpha^{\prime}}$ and $\mu $ are
calculated from the following equations:
\begin{equation}\label{5}
\langle S_{i,\alpha}S_{i,\alpha^{\prime}}\rangle =\langle
S_{0,\alpha}S_{0,\alpha^{\prime}}\rangle\;\; \end{equation} with
$i=1\cdots 2d$, where \begin{equation}\label{6}
\langle\cdots\rangle=\frac{\text{Tr}\exp\left(-\beta
H_{\text{eff}}\right)\cdots}{\text{Tr}\exp\left(-\beta H_{\text{eff}}\right)}   
\;\;\end{equation} with \begin{equation}\label{7}
H_{\text{eff}}=\frac{-1}{\beta}\ln\left[\exp\left(\beta\sum_{i,\alpha}J_{0,i}
S_{0,\alpha}S_{i,\alpha}\right)\right]_{\text{av}}-\frac{\beta
J^2}{2}\mu\sum_{i,\alpha}S_{i,\alpha}^2-\frac{\beta
J^2}{2}\sum_{\alpha\neq
\alpha^{\prime}}\sum_{i=1}^{2d}\mu_{\alpha,\alpha^{\prime}}S_{i,\alpha}S_{i,\alpha
^{\prime}}\;\;. \end{equation} 

Above and at the critical point the
spin glass parameters $\langle S_{k,\alpha}S_{k,\alpha^{\prime}}\rangle=
 q_{\alpha,\alpha^{\prime}}=0$ for $\alpha\neq\alpha^{\prime}$ (but
not for $\alpha=\alpha^{\prime}$). It is easy to calculate that the
effective couplings $\mu_{\alpha\neq\alpha^{\prime}}$ obey the same 
conditions. If we are interested in the calculation of the critical
temperature it is sufficient to formulate the equation (\ref{5}) to
the lowest order in $\mu_{\alpha\neq\alpha^{prime}}$. After
straighforward algebra we get for $\alpha\neq\alpha^{\prime}$ the
following result:         
\begin{equation}\label{8} \langle
S_{i,\alpha}S_{i,\alpha^{\prime}}\rangle \approx\beta^2 J^2\mu_{\alpha,\alpha
^{\prime}}\sum_{j=1}^{2d}\left[\langle S_iS_j
\rangle_0\right]_{\text{av}}\;\; \end{equation} with $i,j=1\cdots 2d $
and \begin{equation}\label{9} \langle
S_{0,\alpha}S_{0,\alpha^{\prime}}\rangle \approx\beta^2 J^2\mu_{\alpha,
\alpha^{\prime}}\sum_{i=1}^{2d}\left[\langle S_0S_j
\rangle_0\right]_{\text{av}}\;\;, \end{equation} where $\alpha\neq\alpha^
{\prime}$ and 
\begin{eqnarray} \label{10}
\langle\cdots\rangle_0& =&\frac{\text{Tr}\exp\left(\beta\sum_{i=1}^{2d}
J_{0,i}S_0 S_i+\frac{\beta^2 J^2}{2}\mu\sum_{i=1}^{2d}S_i^2\right)\cdots}
 {\text{Tr}\exp\left(\beta\sum_{i=1}^{2d}
J_{0,i}S_0 S_i+\frac{\beta^2 J^2}{2}\mu\sum_{i=1}^{2d}S_i^2
\right]}\\ \nonumber &= & \frac{\int_{-\infty}^{\infty}\prod_{i=1}^{2d}{\cal D}x_i
\text{Tr}\exp\left[\beta\sum_{i=1}^{2d}\left(J_{0,i}S_0+J\mu^{1/2}x_i\right)S_i
\right]\cdots}{\int_{-\infty}^{\infty}\prod_{i=1}^{2d}{\cal D}x_i
\text{Tr}\exp\left[\beta\sum_{i=1}^{2d}\left(J_{0,i}S_0+J\mu^{1/2}x_i\right)S_i
\right]}\;\;\end{eqnarray} with \begin{equation}\label{11} {\cal D}x_i=\frac{
1}{\sqrt{2\pi}}\exp\left(-x_i^2/2\right)dx_i\;\;. \end{equation}
Thus due to the translational symmetry for averaged over disorder
correlation functions the equation for the critical temperature takes
the form:
\begin{equation}\label{12} \left[\langle S_i^2\rangle_0^2\right]_{\text{av}}+(2d-1)\left[
\langle S_iS_j\rangle_0^2\right]_{\text{av}}=2d\left[\langle S_0S_i\rangle_0
^2\right]_{\text{av}}\;\;, \end{equation} where $i\neq j$ numbers of
arbitrary lateral spins of the cluster. Additionally we must to take
into account the equation: \begin{equation}\label{13} \left[\langle S_i^2
\rangle_0^2\right]_{\text{av}}=\left[\langle
S_0^2\rangle_0^2\right]_{\text{av}} \;\;.\end{equation} 
With Eqs (\ref{9}) and (\ref{10}) after detailed calculations we
can formulate the equation for the critical temperature in terms of
functions $F_0$ to $F_4$ which depend on the temperature, dimension of
the system, and the spin number. The sample averaged correlation function with 
$i\neq j$ has the form \begin{equation}\label{14} \left[\langle
S_iS_j\rangle^2_0\right]_{\text{av}}=\frac{F_1^2}{F_0^2}\;\;,\end{equation} 
whereas \begin{equation}\label{15} \left[\langle S_0S_i\rangle^2_0\right]_
{\text{av}}=\frac{F_3^2}{F_0^2}\;\;, \end{equation} \begin{equation}\label{16}
\left[\langle S_i^2\rangle_0^2\right]_{\text{av}}=\frac{F_2^2}{F_0^2}\;\;,
\end{equation}and \begin{equation}\label{17} \left[\langle S_0^2
\rangle_0^2\right]_{\text{av}}=\frac{F_4^2}{F_0^2}\;\;.\end{equation}

The form of funcions $F_l$ ($l=0\cdots 4$) is following \begin{equation}
\label{18} F_0=\sum_{M=-S}^S\left\{\int_{-\infty}^{\infty}{\cal D}xQ_S\left[
\beta JS\left(M+\mu^{1/2}x\right)\right]\right\}^{2d}\;\;,\end{equation} 
 \begin{eqnarray}\label{19} F_1 &=& S^2\sum_{M=-S}^S
\left\{\int_{-\infty}^{\infty}{\cal D}xB_S\left[\beta JS\left(M+\mu^{1/2}x
\right)\right]Q_S\left[\beta
JS\left(M+\mu^{1/2}x\right)\right]\right\}^2 \\ \nonumber  &\times &\left\{\int_{-\infty}^{\infty}{\cal D}xQ_S\left[\beta JS\left(M+\mu^{1/2}x\right)
\right]\right\}^{2d-2}\;\;,\end{eqnarray}
\begin{eqnarray}\label{20}
F_2 &=&S
\sum_{M=-S}^S \Biggl(\int_{-\infty}^{\infty}{\cal D}x\left\{B_S^{\prime}\left[
\beta JS\left(M+\mu^{1/2}x\right)\right]+B_S^2\left[\beta
JS\left(M+\mu^{1/2}x\right)\right]\right\}\\ \nonumber &\times & Q_S\left[
\beta JS\left(M+\mu^{1/2}x
\right)\right]\Biggr)\\ \nonumber  &\times &\left\{\int_{-\infty}^{\infty}{\cal D}x
Q_S\left[\beta JS\left(M+\mu^{1/2}x\right)\right]\right\}^{2d-1}\;\;, 
\end{eqnarray} \begin{eqnarray}\label{21} F_3 &= &S\sum_{M=-S}^S\Biggr( M\left\{\int_{
-\infty}^{\infty}{\cal D}xB_S\left[\beta JS\left(M+\mu^{1/2}x
\right)\right]Q_S\left[\beta JS\left(M+\mu^{1/2}x\right)\right]\right\} \\
\nonumber &\times &\left\{\int_{-\infty}^{\infty}{\cal D}xQ_S\left[\beta JS\left(M+\mu^{
1/2}x\right)\right]\right\}^{2d-1}\Biggl)\;\;,\end{eqnarray} and 
\begin{equation} \label{22} F_4=\sum_{M=-S}^S M^2\left\{\int_{
-\infty}^{\infty}{\cal D}xQ_S\left[\beta JS\left(M+\mu^{
1/2}x\right)\right]\right\}^{2d}\;\;.\end{equation} In Eqs
(\ref{16}- \ref{20}) \begin{equation}\label{23}
Q_S\left(y\right)=\frac{\sinh\left[y\left(1+\frac{1}{2S}\right)\right]}{\sinh
\left(\frac{y}{2S}\right)}\;\;,\end{equation}
\begin{equation}\label{24} B_S\left(y\right)= \left(1+\frac{1}{2S}\right)\coth
\left[\left(1+\frac{1}{2S}\right)y\right]-\frac{1}{2S}\coth\left(\frac{y}{2S}
\right)\;\;,\end{equation} is the Brillouin function, and \begin{equation}
\label{25} B_S^{\prime}\left(y\right)=\frac{dB_S^{\prime}\left(y\right)}{dy}
\;\;.\end{equation} Taking into account Eqs (\ref{12}) and (\ref{13})
together with (\ref{14})- (\ref{17}) we can write equations for the
critical temperature as follows:
\begin{equation}\label{26} \left(2d-1\right)F_1^2+F_2^2=2dF_3^2\;\; 
\end{equation} and \begin{equation}\label{27}F_2=F_4\;\;.\end{equation}
Obviously the solution of (\ref{26}) and (\ref{27}) needs numerical
calculations.

Which concerns the linear susceptibility in zero magnetic field we define 
it as follows:
\begin{equation}\label{28} \chi=\frac{1}{N}\sum_i^N\left[\left(\frac{d\langle
S_i\rangle_{T,h}}{dh}\right)_{|h=0}\right]_{\text{av}}\;\;, \end{equation}
where $\langle\cdots\rangle_{T,h}$ denotes the thermal averaging
with the  Hamiltonian (\ref{1}), when the term $-h\sum_i^NS_i$ is
added. A first step is to calculate the susceptibility in the
paramagnetic phase with local magnetizations $\langle
S_i\rangle_T=0$, where $\langle S_i\rangle_T=\langle S_i\rangle_{T,h=0}$.
In that case \begin{equation}\label{29} \chi=\frac{\beta}{N}\sum_{i,j}^N\left[
\langle S_iS_j\rangle_T\right]_{\text{av}}\;\;.\end{equation} It is
easy to show that for $h=0$ and the symmetric probability distribution
for $J_{i,j}$ \begin{equation}\label{30}\left[
\langle S_iS_j\rangle_T\right]_{\text{av}}=\delta_{i,j}\left[
\langle S_i^2\rangle_T\right]_{\text{av}}\;\;.\end{equation}
After
some calculation (see, for example Ref.\onlinecite{8}) we get that
\begin{equation}\label{31} \chi=\beta\left[\langle
S_k^2\rangle_0\right]_{\text{av}}\;\; \end{equation} with $k=0\cdots 2d$.
\section{Results}\label{sec:res}  Our results are
illustrated by plots \ref{fig1}- \ref{fig4}. In Fig. \ref{1} the
dependence of the citical temperature $T_c$ (in units of the constant
$J$) scaled by $\sqrt{2d}$ of the dimension of the system $d$ 
for a few values of spin $S=1/2, 1,3/2, 2$ and $3$ are given. The
larger the spin, the higher the corresponding line. In Fig. \ref{2} 
variations of the $T_c/\sqrt{2d}$ with spin number for $d=2$ ( the
lower line) and $d=3$ (the upper line) are plotted. It is seen that
the critical tempetature, in general, increases with the increasing of the spin 
number but this dependendence cannot be represented in an explicit
form and is more nontrivial comparing to the simple magnetic systems
as, for example, for a ferromagnet where $T_c\sim S$. The reason about
scaling of the ctitical temperature by $\sqrt{2d}$ will be explained in
the next Section. On Figs \ref{fig3} and \ref{fig4} we show the
dependence of the linear susceptibility in the paramagnetic phase of
the temperature $\left(J\equiv 1\right)$ for $d=2$ and $d=3$, respectively. 
The values of $S$ are $1, 3/2$ and $2$. The larger the spin the
higher the corresponding line. Obviously the lines in Figs
\ref{fig3} and \ref{fig4} terminate at the critical temperature
since to calculate $\chi$ below $T_c$ we must enter into theory the
spin glass order parameters. At present our purpose is only to show
that the linear susceptibility, even in the paramagnetic phase, is a
nontrivial function of the temperature for spins higher than $S=1/2$.    
From plots \ref{fig3} and \ref{fig4} it is seen a tendency that
$\chi$ increases with increasing of $S$. 
\section{Final remarks}\label{sec:final}The studies of the spin
glasses with the short ranged interaction is undoubtetly a difficult
problem among the theories of amorphous systems since the complicated
nature of the randomness interplays spatial correlations of spins. At
the present practically there is no developed a systematic method to
investigate such systems, as in the case of the long ranged (more
strictly infinite- ranged) Sherrington- Kirpatrick type models where
MFA is valid \cite{1a,1}.
It is expected that the BPA is able to give more
aacurate estimation of the critical temperature for the spin glass
systems with short range iteraction than the MFA. Which concerns our
problem a natural question arises which is result when the
dimension of the system is ifinite. It can be easy shown that if
$d\rightarrow \infty$ one obtains the Sherrington- Kirpatrick theory
for the Ising spin glass with an arbitrary spin. After rescaling
$J_{i,j}\rightarrow J_{i,j}/\sqrt{2d}$, $J\rightarrow J/\sqrt{2d}$
and changing
$\mu_{\alpha,\alpha^{\prime}}=2dq_{\alpha\neq\alpha^{\prime}}$,
$\mu=2dp$ with $p=q_{\alpha,\alpha}$ proceeding in the similar line
as in Ref. \onlinecite{5} one obtains \begin{equation}\label{32} 
q_{\alpha\neq\alpha^{\prime}}=\langle
S_{\alpha}S_{\alpha^{\prime}}\rangle\end{equation} and 
\begin{equation}\label{33} p= \langle S_\alpha^2\rangle\;\;,\end{equation}
where \begin{equation}\label{34}\langle\cdots\rangle=
\frac{\text{Tr}\exp\left(\frac{\beta^2J^2}{2}\sum_{\alpha\neq\alpha^{\prime}}
q_{\alpha,\alpha^{\prime}}S_{\alpha}S_{\alpha^{\prime}}+p\sum_{\alpha}\S_\alpha
^2\right)\cdots}{\text{Tr}\exp\left(\frac{\beta^2J^2}{2}\sum_{\alpha\neq\alpha
^{\prime}}q_{\alpha,\alpha^{\prime}}S_{\alpha}S_{\alpha^{\prime}}+p\sum_{
\alpha}S_{\alpha}^2\right)}+{\cal O}\left(d^{-1/2}\right)\;\;, \end{equation}
where $S_\alpha$ is the $z$- component of the spin operator referred
to an arbitrary site. Hence in Figs \ref{fig1} and \ref{fig2} in order to 
obtain resonable results for higher spins we scaled the critical
temperature, expressed in units $J$, by $\sqrt{2d}$. We are aware
that our consideration is a first step to recognize some properties
of the Ising SG with an arbitrary spin. It would be interesting to
obtain the properties of the system in the SG phase at least in the
replica symmetric theory. This is a complicated task even for
$S=1/2$. Therefore, further work will be necessary to elucidate this problem.  
\section*{Acknowledments} 
 We acknowledge the financial support ofthe Polish Committee for Scientific
 Research (K. B. N.), Grant No 2 P03B 034 11 . 

\begin{figure}
\caption{ The variation of the critical temperature  $T$  rescaled by the
factor $\left(2d\right)^{-1/2}$ with the dimension $d$ of the system
for spin numbers $S=1/2, 1, 3/2, 2$ and $3$  . The larger the spin the higher 
the corresponding line. Here $J\equiv 1$.}  \label{fig1} 
 \end{figure}
\begin{figure} 
\caption{Rescaling critical temperature $T_c/\left(2d\right)^{-1/2}$
 versus the spin $S$ for $d=2$ and $d=3$ marked by  lower and upper
lines, respectively. Here $J\equiv 1 $.}\label{fig2}
\end{figure}
\begin{figure}
\caption{Linear susceptibility $\chi$ in the paramagnetic state as a
function of the temperature in units of $J$ for $d=2$ and $S=1,  3/2$
and $2$. The larger the spin the higher corresponding line. }\label{fig3}
\end{figure}
\begin{figure}
\caption{Linear susceptibility $\chi$ in the paramagnetic state as a
function of the temperature in units of $J$ for $d=3$ and $S=1,  3/2$
and $2$. The larger the spin the higher corresponding line.}\label{fig4}
\end{figure}
\end{document}